\documentclass[twocolumn,preprintnumbers,amsmath,amssymb,showpacs]{revtex4}

\pdfoutput=1 

\usepackage{graphicx}
\usepackage{bm}
\usepackage{dcolumn}
\usepackage{amssymb}
\usepackage{amsmath}
\usepackage{amsfonts}
\usepackage{color}
\usepackage{epsfig}
\usepackage{subfigure}
\newcommand{\ket}[1]{|#1\rangle}
\newcommand{\bra}[1]{\langle #1|}

\begin{document}
\title{
Density Waves Instability and a Skyrmion Lattice on the Surface of Strong
Topological Insulators}
\author{
    Yuval Baum and Ady Stern }
\affiliation{
     Department of Condensed Matter Physics, Weizmann Institute of Science, Rehovot 76100, Israel}

\begin{abstract}
In this work we analyze the instability conditions for spin-density-waves (SDW) formation on the surface of strong
topological insulators. We find that for a certain range of Fermi-energies and strength of interactions the SDW state is favored compared to the unmagnetized and the uniform-magnetization states. We also find that the SDW are of spiral nature and for a certain range of parameters a Skyrmion-lattice may form on the surface. We show that this phase may have a non trivial Chern-number even in the absence of an external magnetic field.
\end{abstract}
\maketitle
The gapless surface states of strong topological insulators have drawn a great deal of attention over the past few years.
These metallic states arise from a topologically non-trivial band structure \cite{FKM,moore} and are protected as long as time reversal symmetry is maintained.
This topological feature may lead to novel quantum states on the surface of these materials, such as Majorana fermions and fractional excitations \cite{majurana,fqh,fqh1,fqh2}.

In a previous work \cite{Baum} it was shown that for a strong enough electron-electron interaction, both of the short range and the Coulomb types, the surface of a strong topological insulator is unstable to the formation of spontaneous uniform magnetization.
Other works \cite{Fu,Jian} suggested that a spin-density-wave state (SDW) is likely to occur due to a strong Fermi-surface nesting.
In this work we explore the conditions under which the SDW state is favored over the unmagnetized and the uniform-magnetization states.
Moreover, we find that for a certain range of parameters, a Skyrmion-lattice may form on the surface. We also elaborate on the topological properties of this phase and claim that a network of one-dimensional chiral channels can be established on the surface.

We start by analyzing the spin-susceptibility function.
The non-interacting Hamiltonian of a $3D$ topological insulator surface can be approximated by the Fu model \cite{Fu}:
\begin{equation} \label{Eq:Fu_H}
\hat{H}=v_0(k_x\sigma_y-k_y\sigma_x)+\frac{\lambda}{2}(k_+^3+k_-^3)\sigma_z
\end{equation}
where $k_\pm=k_x\pm ik_y$, $\sigma$ are the Pauli matrices, $v_0$ is the electron velocity near the Dirac point and $\lambda$ is the warping parameter which originated from the cubic Dresselhaus spin-orbit coupling of the bulk.
Eq.~(\ref{Eq:Fu_H}) introduces an energy scale, $E^*=(\lambda^{-1}v_0^3)^{1/2}$. The Fermi-surface, arising from Eq.~(\ref{Eq:Fu_H}), can be classified to three regions \cite{Chen,Hsieh,Hsieh2}. A circular region for $|E_F|<0.55E^*$, a hexagonal region for $0.55E^*<|E_F|<0.9E^*$ and a 'snowflake' region for $|E_F|>0.9E^*$. Usually, the 'snowflake' region coexists with the bulk states and therefore it is less interesting. Hence, we assume that the Fermi-energy lies in the circular or the hexagonal regions.
 
We denote the energies and normalized eigenstates of Eq.~(\ref{Eq:Fu_H}) as $\epsilon_{\textbf{k},\pm}$ and $U_{\textbf{k},\pm}$.
The stationary ($\omega=0$) and non-interacting spin-susceptibility function is given by:
\begin{equation} \label{Eq:sus}
\chi_0^{\mu\nu}(\textbf{q},T)=\sum_{s,s'=\pm}\sum_\textbf{k}\frac{n_F(\epsilon_{\textbf{k},s})-n_F(\epsilon_{\textbf{k+q},s'})}{\epsilon_{\textbf{k},s}-\epsilon_{\textbf{k+q},s'}}Y_{ss'}^\mu(Y_{s's}^{\nu})^*
\end{equation}
where $n_F$ is the Fermi-Dirac function at temperature $T$ with Fermi-energy $\epsilon_F$ and
$Y_{ss'}^\mu(\textbf{k},\textbf{q})\equiv\langle U_{\textbf{k},s}|\sigma^\mu| U_{\textbf{k+q},s'}\rangle$.

\begin{figure}[b]
\centering
\subfigure[]{
   \includegraphics[scale =0.21] {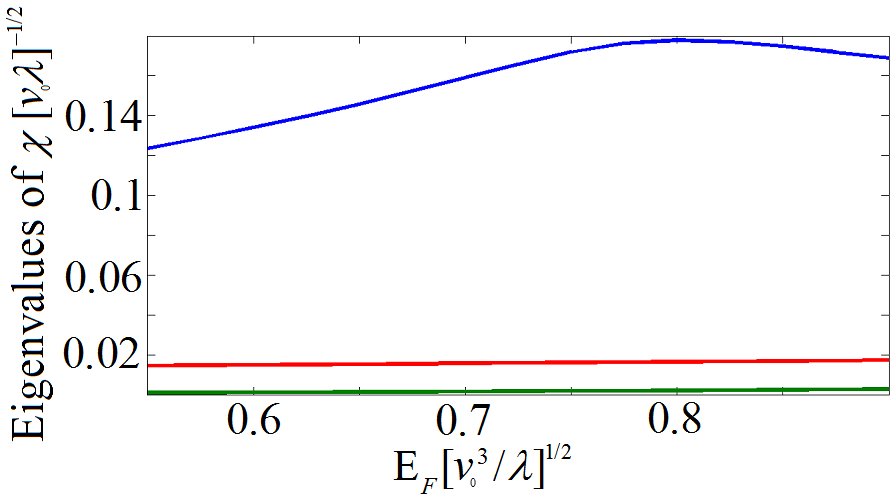}
   \label{Fig1}
 }
\subfigure[]{
   \includegraphics[scale =0.21] {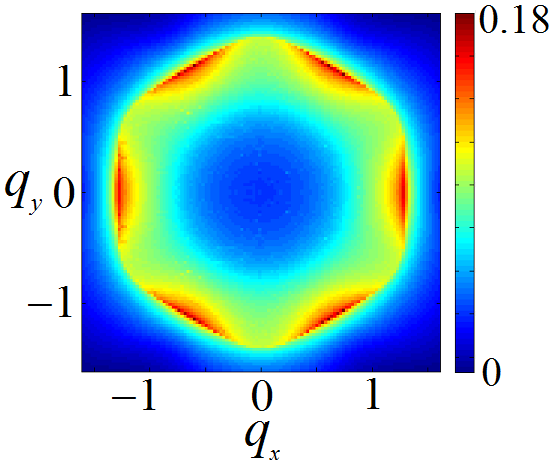}
   \label{Fig2}
 }
\caption{(a) The zero temperature eigenvalues of $\chi_0^{\mu\nu}$ for a fixed $\textbf{q}=(2k_F,0)$ as a function of the Fermi-energy, where the Fermi-energy is           measured in terms of
         $E^*=(\lambda^{-1}v_0^3)^{1/2}$ and the susceptibility is measured in terms of $(\lambda v_0)^{-1/2}$.
         As the Fermi-energy increases, one of the eigenvalues becomes much larger than the other two. (b) The maximal eigenvalue of $\chi^{\mu\nu}$ as a                   function of $\textbf{q}$, for a fixed Fermi-energy ($E_F=0.7E^*$), where $\textbf{q}$ is measured in terms of $(\lambda^{-1}v_0)^{1/2}$.
         The susceptibility is maximized at the nesting vectors of the Fermi-surface. }
\end{figure}

\begin{figure}[b]
\begin{center}
\includegraphics[scale =0.32]{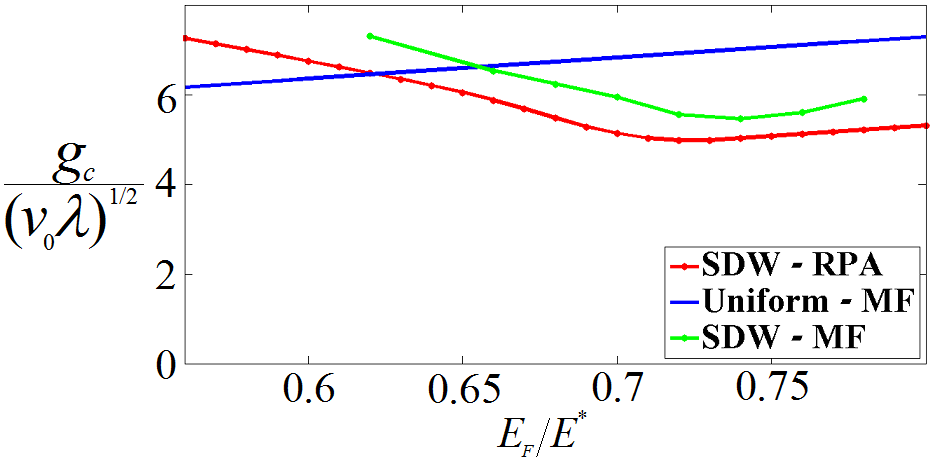}
\caption{ \label{Fig3} %
The critical interaction to have a SDW/uniform-magnetization ground state as a
function of the Fermi-energy at zero temperature. For a certain range of energies the SDW instability is favored compared to the uniform one. }
\end{center}
\end{figure}

The eigenvalues of $\chi_0^{\mu\nu}(\textbf{q},T=0)$, for $\textbf{q}=(2k_F,0)$, as a function of the Fermi-energy are presented in Fig.~(\ref{Fig1}).
Clearly, as the Fermi energy increases, one of the eigenvalues becomes much larger than the other two.
The maximal eigenvalue as a function of $\textbf{q}$ , for a fixed Fermi-energy in the hexagonal region, is presented in Fig.~(\ref{Fig2}). The susceptibility is maximized at the nesting vectors of the Fermi-surface in the hexagonal region.
By analyzing the susceptibility as a function of the temperature (see supplementary), we find that the susceptibility increases dramatically as the temperature decreases. However, it remains finite for all temperatures. This feature arises from the fact that the Fermi-surface is not perfectly nested, otherwise we would expect the susceptibility to diverge at zero temperature. 
The Landau free-energy up to a second order, with the magnetization, $\textbf{m}_\textbf{q}$, as an order parameter, is given by:
\begin{equation}\label{Eq:FL}
F_L=\sum_\textbf{q}m^\mu_\textbf{-q}\chi^{\mu\nu}_\textbf{q}m^\nu_\textbf{q}
\end{equation}
Without interactions, as shown in Fig.~(\ref{Fig1}), the eigenvalues of the susceptibility are non-negative. Hence, Eq.~(\ref{Eq:FL}) is minimized by $\textbf{m}_\textbf{q}=0$ for every $\textbf{q}$.
However, in the presence of a contact interaction, the susceptibility can be expressed in terms of the non-interacting susceptibility using the random phase approximation (RPA), $\chi^{\mu\nu}=\chi_0^{\mu\rho}[(1-g\chi_0)^{-1}]^{\rho\nu}$, where $g$ is the contact-interaction strength. Since the non-interacting susceptibility is finite for all temperatures, we conclude that a critical interaction at which one of the eigenvalues of the susceptibility becomes negative exists for all temperatures. Above that critical value a second order phase transition to a SDW ground state occurs.
The zero-temperature critical interaction for a formation of a SDW ground state as a function of the Fermi-energy, as calculated in the RPA, appears in Fig.~(\ref{Fig3}). The critical interaction (mean-field) to have a uniform magnetization ground state as calculated in \cite{Baum} also appears in Fig.~(\ref{Fig3}).
At the circular region, where the Fermi-surface nesting is negligible, the uniform state is favored. As the Fermi-energy increases to the hexagonal range, the nesting becomes stronger and the SDW state is favored compared to the uniform one.
We also find that the values of the critical interaction for a formation of a SDW ground state are quite sensitive to temperature. The zero temperature values in Fig.~(\ref{Fig3}) are doubled at $T\approx E^*/10$.
 
Above the critical interaction a non-uniform magnetization will occur on the surface. By denoting the normalized eigenvectors of the susceptibility as $\hat{n}_\textbf{q}$, the magnetization becomes
\begin{equation} \nonumber
\textbf{m}(\textbf{r})\propto\sum_{\textbf{q}} \mbox{Re}(\hat{n}_{\textbf{q}})\cos{(\textbf{q}\cdot \textbf{r}+\varphi_\textbf{q})}+\mbox{Im}(\hat{n}_{\textbf{q}})\sin{(\textbf{q}\cdot \textbf{r}+\varphi_\textbf{q})}
\end{equation}

Since all the eigenvectors of the susceptibility contain both real and imaginary parts, each $\textbf{q}$ contributes to the magnetization a spiral SDW propagating in the $\textbf{q}$ direction. The magnitude of the SDW is almost constant and its direction varies continuously in space.
The relative phase between each pair of SDW, $\Delta\varphi_\textbf{q}$, is physical. Only two of these relative phases are independent, and choosing them is equivalent to choosing the origin of the SDW. This freedom will be manifested as Goldstone modes (phonons).

Near the phase transition, only the nesting vectors $\{\textbf{Q}_i\}$ will contribute to the magnetization. However, these SDW do not necessarily coexist.
The Landau free-energy up to a fourth order is:
\begin{equation} \label{Eq:FL4}
F_L\sim\sum_{i}\chi_{_{\textbf{Q}_i}}|\textbf{m}_{_{\textbf{Q}_i}}|^2+U|\textbf{m}_{_{\textbf{Q}_i}}|^4+V|\textbf{m}_{_{\textbf{Q}_i}}|^2
|\textbf{m}_{_{\textbf{Q}_{i+1}}}|^2
\end{equation}
where $U$ and $V$ are four-spin correlation functions. Clearly, by tuning the ratio between $U$ and $V$, the system can change its phase from a phase where all the spin-spirals coexist, to a stripes phase where only one of the spin-spiral has a magnitude that is different from zero. 
In general, $U$ and $V$ depend on the dimensionless parameter $x\equiv g(\lambda v_0)^{-1/2}$ and on the dimensionless Fermi-energy ($E_F/E^*$).
Though the exact calculation of $U$ and $V$ is difficult, a numerical mean-field calculation suggests that both phases may occur for different combinations of these parameters.  

To the end of this paper, we will focus on the situation where the ratio between $U$ and $V$ is such that the favored ground state is the coexisting phase.
The superposition of the non-coplanar spin-spirals form then a triangular Skyrmion-lattice on the surface \cite{sky1,sky2}.

\begin{figure}[b]
\begin{center}
\includegraphics[scale =0.45]{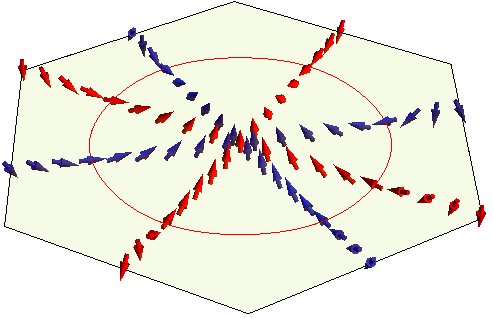}
\caption{ \label{Fig4} %
The direction of the magnetization at selected points in the unit cell of the triangular lattice for the $Q_{top}=1$ phase. In the core, the magnetization is in the direction normal to the surface. The red line denotes all the points where the magnetization is in the plane. }
\end{center}
\end{figure}

The components of the magnetization form a periodic structure with a triangular symmetry. Each unit cell of the triangular lattice holds a Skyrmion with a unit Pontryagin-number \cite{sky3}, $Q_{top}=\pm1$, and a zero net magnetization. In the core of each Skyrmion the magnetization points in the direction normal to the surface and it rotates continuously towards the opposite direction at the edges of the unit cell. An illustration of the magnetization direction in a unit cell appears in Fig.~(\ref{Fig4}). Clearly, the system chooses spontaneously one of the two time-reversal-breaking states which are characterized by opposite Pontryagin-number, $Q_{top}=\pm1$.    

In the $\nu=1$ quantum Hall state \cite{QHsky1,QHsky2}, the Skyrmion-lattice phase breaks both the translation and the $U(1)$ symmetry of the single-Skyrmion state. Unlike the quantum Hall case, in this case there is no rotational symmetry and therefore we expect that the only massless Goldstone modes will be the standard phonons of a two dimensional triangular lattice. 

\begin{figure}[t]
\begin{center}
\includegraphics[scale =0.34]{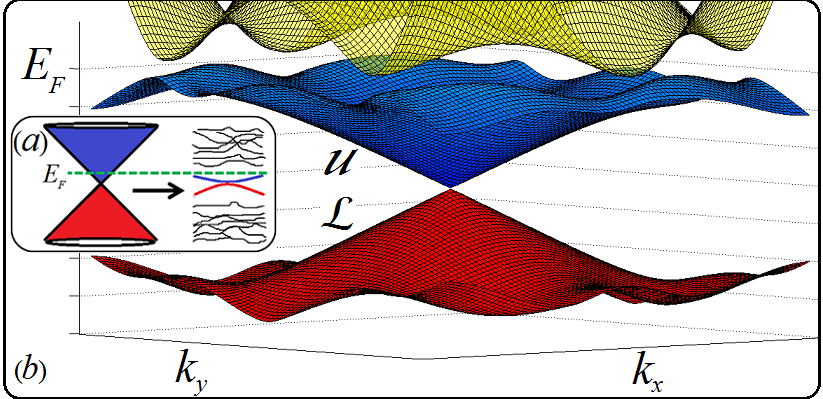}
\caption{ \label{Fig5} %
(a) An illustration of the transition from a continuum Dirac cone to a periodic band structure. (b) Magnification of the surface band structure near the Dirac-point for $E_F=0.7E^*$ and $g=1.1g_c$, where $g_c$ is the critical interaction strength. Where $k_x$ and $k_y$ are in the first Brillouin zone and measured in terms of $(\lambda^{-1}v_0)^{1/2}$ and the energy is measured in terms of $E^*=(\lambda^{-1}v_0^3)^{1/2}$. A gap is opened near the Fermi-energy at every point in k-space.}
\end{center}
\end{figure}

Motivated by the RPA results, we now discuss the mean-field (MF) theory of the Skyrmion-lattice phase. This mean-field treatment will allow us a better understanding of the Skyrmion-lattice phase and its topological features.
We consider Eq.~(\ref{Eq:Fu_H}) as the non-interacting part of the Hamiltonian. We also assume a non-vanishing expectation value of the magnetization:
\begin{equation} \nonumber
\langle\psi^\dagger(\textbf{r})\,\boldsymbol{\sigma}\,\psi(\textbf{r})\rangle=\textbf{m}(\textbf{r})=\sum_\textbf{Q}\textbf{m}_{_\textbf{Q}}e^{i\textbf{Q}\cdot \textbf{r}}
\end{equation}
where $\textbf{m}(\textbf{r})$ describes the Skyrmion-lattice that we found in the RPA, and $\textbf{Q}$ are the nesting vectors of the Fermi-surface in the hexagonal region. 
In the mean-field approximation, the interaction part of the Hamiltonian becomes:
\begin{align} \label{Eq:HI}  
\mathcal{H_{_I}}^{MF}=g\sum_\textbf{Q}|\textbf{m}_{_\textbf{Q}}|^2-2g\sum_{\textbf{Q},\textbf{k}}C^\dagger_\textbf{k}\,( \textbf{m}_{_\textbf{Q}}\cdot \boldsymbol{\sigma})\, C_{\textbf{k-Q}}
\end{align}
where $C_{\textbf{k}}$ are the Fourier components of $\psi(\textbf{r})$ and $g$ is the interaction strength of a contact interaction. The full mean-field Hamiltonian, including the kinetic part Eq.~(\ref{Eq:Fu_H}), can be solved now by using Bloch theorem, to yield a band structure. To the end of this paper, we restrict the Fermi-energy to be at the hexagonal range and we assume that the magnetization energy is much smaller than that scale.
The two continuum branches of the Dirac-cone split into a periodic band structure Fig.~(\ref{Fig5}(a)).
We define the ground-state MF energy:
\begin{equation} \label{Eq:m_f_E}
E_{g.s}^{MF}=\sum_{\textbf{Q}}gm_{_{\textbf{Q}}}^2+\sum_{n,\textbf{k}}\epsilon_{n,\textbf{k}}
\end{equation}
where the summation of $\textbf{k}$ is over the first Brillouin zone and the summation of $n$ is over the occupied bands.
By minimizing Eq.~(\ref{Eq:m_f_E}) with respect to the magnetization, we find that a critical interaction exists, above which there is a non-trivial minimum to  Eq.~(\ref{Eq:m_f_E}). This critical interaction as a function of the Fermi-energy appears in Fig.~(\ref{Fig3}). We notice that the MF values of the critical interaction agree quite well with the RPA results. We also find the phase transition to the gapped phase to be a second order phase transition where the gap grows as $\sqrt{g-g_c}$.
As shown in \cite{Baum}, in the known topological insulators such as Bi$_2$Te$_3$ and Bi$_2$Se$_3$, these values of critical interaction strength are too strong for an instability, albeit not by a large factor, making the instability an issue that may be relevant for other topological insulators.

The spectrum of the surface splits into many bands which, in general, cross each other. However, we found that the two bands near the Dirac-point cross each other only at the Dirac-point and are gapped from all the other bands.
The two bands near the Dirac-point for representative values of the parameters are presented in Fig.~(\ref{Fig5}(b)).
Clearly, while near the Dirac-point the original branches of the Dirac-cone are almost unchanged, at the Fermi-energy the change is dramatic. A gap is opened at the Fermi-energy and the surface spectrum becomes fully gapped.
Moreover, the surface spectrum is not symmetric with respect to the Dirac point, since the magnetization has a non-zero Pontryagin-number (winding) which differentiates between states with different chirality.        

We now turn to explore the topological properties of the surface. In the absence of interactions, the spectrum of the surface is gapless. By adding a mass term, which is equivalent to a uniform magnetization, a gap is opened at the Dirac-point and a half-integer Hall conductivity arises \cite{surf_QH}.
For an infinite Dirac-cone, this half-integer Hall conductivity is precisely quantized. However, in real systems the Dirac bands have an energy cutoff and therefore the half-integer Hall conductivity is not precisely quantized. It deviates from the half-integer value by an amount which scales as the gap energy divided by the cutoff energy. 
Moreover, the gap at the Dirac-point can be opened due to interactions or by applying an external magnetic field. In both cases the gap that is opened is rather small ($\sim1_{meV}$) which makes the observation of the half-integer Hall conductivity a hard task, since experimentally, it is required to fine tune the Fermi-energy to the narrow gap.

As opposed to this narrow gap, the hexagonal range is covering a wide energy window, both in the upper and in the lower branch. In the well known topological insulators this energy window, $\Delta E_{hex}\approx 0.35E^*$, is of the order of $\sim100_{meV}$. 
As shown in Fig.~(\ref{Fig3}), by tuning the Fermi-energy to any value in the hexagonal range and for a strong enough interaction, a gap is opened at the Fermi-energy and a quantum Hall effect may be observed. 

The spectrum of the Skyrmion-lattice phase is always gapped at the Fermi-energy and the lattice symmetry of this phase yields a Brillouin zone and a wave-function of the form of a Bloch wave-function. This allows us to define a Chern-number associated with each band:
\begin{equation} \label{Eq:chern}
C_j=\frac{i}{2\pi}\sum_{\textbf{k}}Tr(P_j[\partial_{k_x}P_j,\partial_{k_y}P_j])
\end{equation}
where, $P_j$ is the projector to the $j^{th}$ band.
We denote the two bands near the Dirac-point as $\mathcal{L}$ and $\mathcal{U}$, where $\mathcal{L}$ is the lower band and $\mathcal{U}$ is the upper band (see Fig.~(\ref{Fig5})). Of course, there are many more bands above and below these two, both surface and bulk bands. However, the $\mathcal{L},\mathcal{U}$  bands are gapped from all other bands and we will treat them as an isolated band inside the bulk gap. Using Eq.~(\ref{Eq:chern}), we find that  when the Fermi-energy is at the hexagonal range the Chern-number of this band is $C_{\mathcal{L}+\mathcal{U}}=Q_{top}=\pm1$ (see supplementary). This Chern-number of the combined band is independent of the existence of a mass term that introduces an energy gap between them. In the presence of a mass term these two bands are separated and their total Chern-number is unchanged.

\begin{figure}[t]
\begin{center}
\includegraphics[scale =0.31]{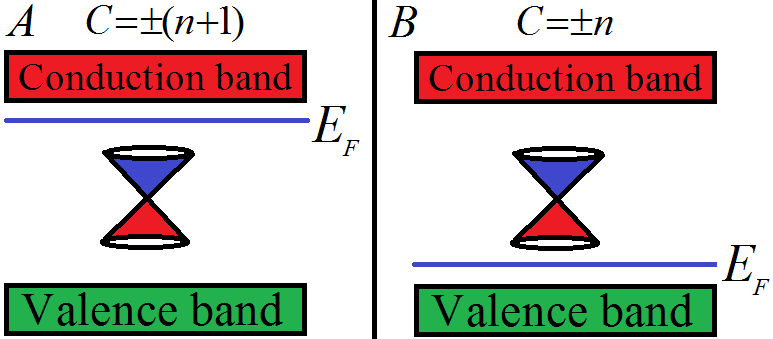}
\caption{ \label{Fig8} %
An illustration of the surface band structure in the Skyrmion-lattice phase. The valence and conduction bands are composed of bulk and surface bands.
The bands $\mathcal{L}$ and $\mathcal{U}$ are separated from the other bands by an energy gap. The Chern-numbers of the two scenarios differ by an odd integer. }
\end{center}
\end{figure}

It is instructive to examine how transport on the surface of a thick slab depends on the Fermi-energy when electron-electron interaction is strong enough for spontaneous magnetization to occur. When the Fermi-energy is tuned to the hexagonal range in the lower branch, a gap is opened beneath the $\mathcal{L}$ band and the occupied bands below the Fermi energy contribute a Chern-number $\pm n$. The $\pm$ signs refer to the two possible time-reversal-breaking states. In general, the contributions to this $n$ come from the three dimensional bands within the bulk of the slab and the occupied bands on its surfaces. However, since the breaking of time-reversal-symmetry is limited to the surface, the bulk bands do not contribute. As the Fermi-energy is increased, the Chern-number does not change as long as the system is within the Skyrmion phase, since the energy gap does not close. Once the Fermi-energy reaches the region where the magnetization becomes uniform, the gap shifts away from the Fermi-energy to the Dirac point, and the Hall conductivity ceases to be quantized. It becomes quantized again when the Fermi-energy is at the Dirac point, in an energy gap. Then, the Hall conductivity of the entire slab is quantized to an integer, and that of each surface is approximately quantized to a half-integer. Finally, when the Fermi-energy is increased further, it passes through a gapless region in the circular range and gets to the hexagonal range on the positive side of the spectrum. 
Again, the system chooses spontaneously between the two time-reversal-breaking states and the occupied bands will contribute a Chern-number of $\pm(n+1)$. The particular value of $n$ is not universal, and would depend on details of the band structure. We can, however, generally say that the difference in Hall conductivity between the Fermi-energy lying in the lower and upper hexagonal ranges is an odd integer. These two cases are illustrated in Fig.~(\ref{Fig8}).

\begin{figure}[b]
\centering
\subfigure[]{
   \includegraphics[scale =0.28] {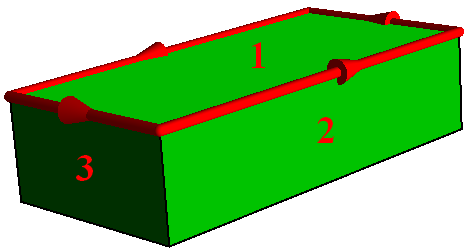}
   \label{Fig6}
 }
\subfigure[]{
   \includegraphics[scale =0.28] {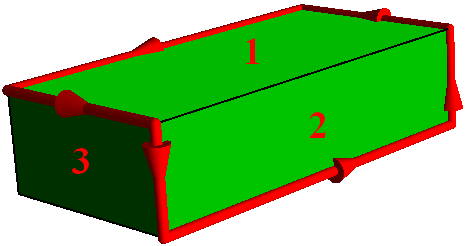}
   \label{Fig7}
 }
\caption{An illustration of the one-dimensional edge channels on the surface of a topological insulator. (a) The Fermi-energy on surface 1 is tuned to the hexagonal range at the upper branch while the Fermi-energies of all the other surfaces are tuned to the hexagonal range at the lower branch. (b) Same scenario as in (a) only now the Fermi-energies on surfaces 1 and 2 are tuned to the hexagonal range at the upper branch. }
\end{figure}

The analysis above points out to two types of situations where the Hall conductivity on a surface may vary in space. First, that may happen when time-reversal is broken in opposite ways in different regions. And second, that may happen when the Fermi-energy varies in space, due to a variation between local density of electrons. In both cases, the domain walls must carry one-dimensional chiral edge channels. Notably, these chiral edge modes occur without an external magnetic field. 
Such one-dimensional edge channels on the surface of topological insulators may arise as a consequence of a magnetic field, as discussed in \cite{1d_channel}. Here we point out that in the presence of strong enough interactions, they may appear without an external field and without the need to fine tune the Fermi-energy. An illustration of this scenario is presented in Fig.~(\ref{Fig6},\ref{Fig7}). The starting point is a topological insulator where the Fermi-energy of all the surfaces is tuned to the hexagonal range at the lower branch. By tuning the Fermi-energy of selected surfaces to the hexagonal range at the upper branch, one-dimensional edge channels will appear. This tuning may be done by an external voltage. 

To summarize, we considered here the possibility that a non-uniform magnetization spontaneously appears on the surface of a three dimensional strong topological insulator due to interactions between surface electrons.
We assumed that the Fermi-energy is such that the Fermi-surface arising from a $2D$ effective Dirac Hamiltonian is strongly nested. We treated interactions within the mean-field and the random phase approximations. We found that for a strong enough interaction the surface may be unstable to the formation of a Skyrmion-lattice phase which is a result of a multiple SDW states.
We found the transition from the unmagnetized to Skyrmion-lattice phase to be of second order as a function of interaction strength.
We also found that as long as the Fermi-energy is located at the hexagonal range, the dependence of the critical interaction strength on Fermi-energy is rather mild while the temperature dependence is rather strong.
We showed that even in the absence of an external magnetic field a non-trivial Chern-number arises from the mean-field Bloch wave function of the surface in the Skyrmion-lattice phase and that this Chern-number can be controlled by tuning the Fermi-energy. Using this feature, we suggested that a network of one-dimensional chiral channels can be established on the surfaces of a strong topological insulator with a strong enough electron-electron interaction. 

The authors thank the US-Israel Binational Science Foundation and the Minerva foundation for financial support.
\renewcommand{\theequation}{S-\arabic{equation}}
\renewcommand{\thefigure}{S-\arabic{figure}}
\section*{Supplementary material}
\subsection*{Temperature Dependence}
\begin{figure}[h]
\begin{center}
\includegraphics[scale =0.32]{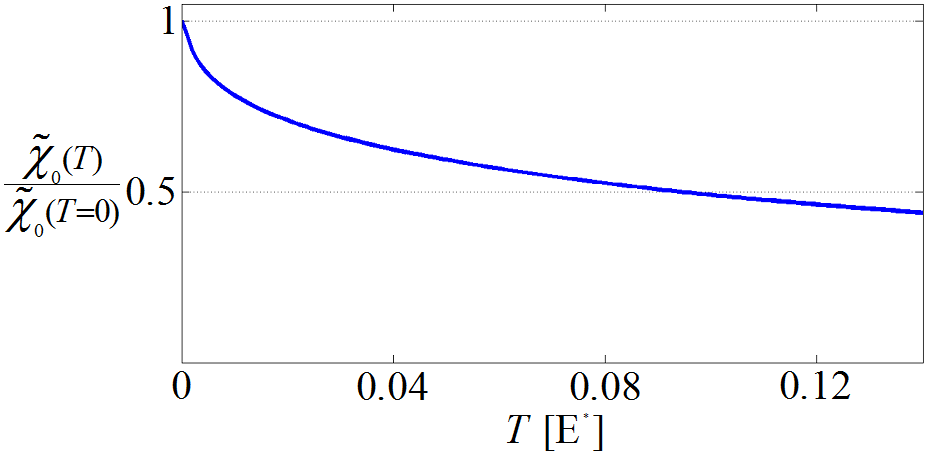}
\caption{ \label{Fig:X_temp} %
The ratio of the largest eigenvalues of $\chi^{\mu\nu}(\textbf{q},T)$ and $\chi^{\mu\nu}(\textbf{q},T=0)$ as a function of the temperature for a fixed $\textbf{q}=(2k_F,0)$. The temperature is measured in terms of $E^*=(v_0^3/\lambda)^{1/2}$. The susceptibility increases as the temperature decreases.}
\end{center}
\end{figure}

The stationary ($\omega=0$) and non-interacting spin-susceptibility function, $\chi_0^{\mu\nu}(\textbf{q},T)$, is given in Eq.~(2).
When the Fermi-energy is at the hexagonal range and $\textbf{q}$ is close to one of nesting vectors, one of the eigenvalues of $\chi_0^{\mu\nu}$ becomes much larger than the other two.
We denote the largest eigenvalue of $\chi_0^{\mu\nu}$ as $\tilde{\chi}_0$. 
We found that $\tilde{\chi}_0$ is maximized at the nesting vectors of the Fermi-surface.
Now, we would like the examine the temperature dependence of $\tilde{\chi}_0$. In order to do so, we set $\textbf{q}=(2k_F,0)$. The ratio of $\tilde{\chi}_0(T)$ and $\tilde{\chi}_0(T=0)$ as a function of the temperature appears in Fig.~(\ref{Fig:X_temp}), where the temperature is measured in terms of $E^*$.
Clearly, the susceptibility increases as the temperature decreases. However, it remains finite for all temperatures.
The critical interaction strength scales as $\tilde{\chi}_0^{-1}$, hence, its zero-temperature values are doubled at $T\approx E^*/10$.
The values of $\tilde{\chi}_0(\textbf{q},T)$ at the other two nesting vectors are identical to Fig.~(\ref{Fig:X_temp}) due to symmetry considerations. 
For other $\textbf{q}$ vectors, the values of $\tilde{\chi}_0(\textbf{q},T)$ are different but the temperature dependence is similar.
\subsection*{Chern-numbers Calculations}
The magnetization of a Skyrmion-lattice is:
\begin{equation} \label{mag}
\textbf{m}(\textbf{r})=M\sum_{\textbf{Q}} \mbox{Re}(\hat{n}_{_{\textbf{Q}}}e^{\textbf{Q}\cdot \textbf{r}+\varphi_{_\textbf{Q}}})
\end{equation}
where the summation is over the nesting vectors $\{\textbf{Q}\}$, $\hat{n}_{_{\textbf{Q}}}$ are the eigenvectors of the spin-susceptibility, $M$ is the total amplitude of the Skyrmion-lattice and $\varphi_{_\textbf{Q}}$ is the phase of each SDW. 
The mean-field Hamiltonian of a strong topological insulator surface with Eq.~(\ref{mag}) as an expectation value of
$\langle\psi^\dagger(\textbf{r})\,\boldsymbol{\sigma}\,\psi(\textbf{r})\rangle$ is:
\begin{align} \label{H_MF} 
\mathcal{H}=g\sum_\textbf{Q}|\textbf{m}_{_\textbf{Q}}|^2+\sum_{\textbf{k}}C^\dagger_\textbf{k}\hat{H}_0 C_{\textbf{k}}-2g\sum_{\textbf{Q},\textbf{k}}C^\dagger_\textbf{k}( \textbf{m}_{_\textbf{Q}}\cdot \boldsymbol{\sigma}) C_{\textbf{k-Q}}
\end{align}
where $C_{\textbf{k}}$ are the Fourier components of $\psi(\textbf{r})$, and $\hat{H}_0$ is the non-interacting part which can be approximated by the Fu model.
Eq.~(\ref{H_MF}) can be solved numerically via Bloch theorem. That introduces a band structure and a set of eigenvectors, $\ket{\psi_{n,\textbf{k}}}$, at each point in the lattice Brillouin zone. The Chern-number calculation can be done numerically by defining the projection matrix, $P_{n,\textbf{k}}=\ket{\psi_{n,\textbf{k}}}\bra{\psi_{n,\textbf{k}}}$, and using:
\begin{equation} \nonumber
C_n=\frac{i}{2\pi}\sum_{\textbf{k}}Tr(P_{n,\textbf{k}}[\partial_{k_x}P_{n,\textbf{k}},\partial_{k_y}P_{n,\textbf{k}}])
\end{equation} 
In order to perform the calculation, we must set the parameters of the problem. In particular, we must choose the exact form of the magnetization by setting 
the amplitude, $M$, and the phases, $\varphi_\textbf{q}$, in Eq.~(\ref{mag}).
We found numerically that for any non-zero value of the amplitude $M$, a gap is opened at the Fermi-energy. Hence, any two states with two different positive (or two different negative) values of $M$ can be connected continuously without closing the gap. Consequently, they must share the same Chern-number. Moreover, two states that differ only in the sign of $M$ are two time-reversal-partners and therefore their Chern-numbers are equal in magnitude and opposite in sign.
Hence, we conclude that the Chern-number of the system can not depend on the magnitude of $M\neq 0$, and that the sign of $M$ affects only the sign of the Chern-number.
For that reason, we can choose a specific value of $M$ without losing the generality. For the actual calculation, we chose $E_F=0.7E^*$ and $g/g_c=1.2$ where $g_c$ is the appropriate critical interaction. We set $M$ to be positive such that $U_m/E_F=0.01$, where $U_m$ is the average magnetic energy per unit-cell:
\begin{equation} \nonumber
U_m=\frac{g}{S_{u.c}}\int_{u.c}d^2\textbf{r}|\textbf{m}(\textbf{r})|^2
\end{equation}
where the integration is over a lattice unit-cell and $S_{u.c}$ is a unit-cell area.

The phases, $\varphi_\textbf{q}$, determine the origin of the Skyrmion-lattice. Clearly, the spectrum and the Chern-numbers are independent of these phases. Therefore, without loss of generality, we set all the phases to be zero. 
After the determination of $E_F$, $g$ and the exact form of $\textbf{m}(\textbf{r})$, the numerical calculation of the Chern-number is a straightforward task.

\end{document}